\documentclass[conference]{IEEEtran}
\usepackage{graphicx}
\usepackage{float}
\usepackage{url}
\usepackage[utf8]{inputenc} 
\usepackage{amsmath}
\usepackage{amsfonts}
\usepackage[usenames,dvipsnames]{color}
\usepackage{natbib}
\begin{document}

\title{POPmine: Tracking Political Opinion on the Web}

\author{\IEEEauthorblockN{Pedro Saleiro}
\IEEEauthorblockA{DEI-FEUP / Sapo Labs UP\\
University of Porto\\
Rua Dr. Roberto Frias, s/n\\
4200-465 Porto\\
 Portugal}
\and
\IEEEauthorblockN{S\'{i}lvio Amir}
\IEEEauthorblockA{IST/INESC-ID\\
Rua Alves Redol, 9\\
1000-029 Lisboa\\
Portugal}
\and
\IEEEauthorblockN{M\'{a}rio Silva}
\IEEEauthorblockA{IST/INESC-ID\\
Rua Alves Redol, 9\\
1000-029 Lisboa\\
Portugal}
\and
\IEEEauthorblockN{Carlos Soares}
\IEEEauthorblockA{DEI-FEUP / INESC-TEC\\
University of Porto\\
Rua Dr. Roberto Frias, s/n\\
4200-465 Porto\\
 Portugal}}

\maketitle

\begin{abstract}
The automatic content analysis of mass media in the social sciences has become necessary and possible with the raise of social media and computational power. One particularly promising avenue of research concerns the use of opinion mining.
We design and implement the POPmine system which is able to collect texts from web-based conventional media (news items in mainstream media sites) and social media (blogs and Twitter) and to process those texts, recognizing topics and political actors, analyzing relevant linguistic units, and generating indicators of both frequency of mention and polarity (positivity/negativity) of mentions to political actors across sources, types of sources, and across time.

\end{abstract}

\section{Introduction}

The driving forces that shape public opinion are a traditional object of study by journalists, politicians and social scientists. However, with the raise of social media, the connections between political and economic events, how people come to realize them and how they react has become fuzzier and complex phenomena. Blogs, Twitter or Facebook provide additional sources of politically relevant information to people in almost real time. On the other hand, people also publish their own reactions and opinions, therefore social media data might provide an approximate picture of mass public opinion itself. 

Understanding this relationship is valuable to several actors, including news publishers, political campaign staff, pollsters or social scientists. 
For instance, political parties monitoring social media data can enhance vote targeting, fund raising or volunteer recruitment. Moreover they can create new approaches to predict trends, poll or electoral results or even the impact of certain events. Though, there are several challenges when trying to collect, process and mine social media and online news data for these purposes \cite{chall1}. Tweets are the most challenging type of text, specially when trying to apply named entity recognition or sentiment analysis, mainly due to the ambiguity, lack of contextualization and informal language (unorthodox capitalization, emoticons, hashtags). 

These tasks are technically complex for most of the people interested in tracking political opinion on the web. For this reason, most research has focused on investigating parts of this problem leading to the development of tools that only address  sub-tasks of this endeavour.  This means that there are no open source integrated platforms to support this work. 

The main contribution of this work is the design and implementation of the POPmine system, an open source platform which can be used and extended by researchers interested in tracking political opinion on the web. POPmine operates either in batch or online mode and is able: to collect texts from web-based conventional media (news items in mainstream media sites) and social media (blogs and Twitter); to process those texts, recognizing topics and political actors; to analyze relevant linguistic units; to generate indicators of both frequency of mention and polarity (positivity/negativity) of mentions to political actors across sources, types of sources, and across time. As a proof of concept we present these indicators in a web application tailored for tracking political opinion in Portugal, the POPSTAR website. The system is available as an open source software package that can be used by other researchers from social sciences but also from any other area that is interested in tracking public opinion on the web.

\section{Related Work}

Content analysis of mass media has an established tradition in the social sciences, particularly in the study of effects of media messages,
encompassing topics as diverse as those addressed in seminal studies of newspaper editorials \cite{lasswell1952}, media agenda-setting
\cite{mccombs1972}, or the uses of political rhetoric \cite{moen1990}, among many others. By 1997, Riffe and Freitag \cite{riffe1997}, 
reported an increase in the use of content analysis in communication research and suggested that digital text and computerized means for its
extraction and analysis would reinforce such trend. Their expectation has been fulfilled: the use of automated content analysis has by now surpassed the use of hand coding \cite{neuendorf2002}. The increase in the digital sources of text, on the one hand, and current advances in
computation power and design, on the other, are making this development both necessary and possible, while also raising awareness about the
inferential pitfalls involved \cite{hopkins2010, chall2}.

One avenue of research that has been explored in recent years concerns the use of social media to predict present and future political events, namely electoral results \cite{Bermingham, pred12, pred6, pred3, livne2011party, tumasjan2010predicting, Gayo-Avello2012, connor2010, Chung2011}. Although there is no consensus about methods and their consistency \cite{Metaxas2011, Gayo-Avello2011}. Gayo-Avello \cite{pred_survey} summarizes the differences between studies conducted so far by stating that they vary about period and method of data collection, data cleansing and pre-processing techniques, prediction approach and performance evaluation. Two main strategies have been used to predict elections: buzz or number of tweets mentioning a given candidate or party and secondly the use of sentiment analysis. Different computational approaches have been explored to process sentiment in text, namely machine learning and linguistic based methods \cite{pang2008, Kouloumpis2011, nakov2013}. In practice, algorithms often combine both strategies. One particular challenge when using sentiment is how to aggregate opinions in a timely fashion that can be fed to the prediction method.

The majority of the work using social media data applied to political scenarios consist in ad hoc studies where researchers collect data from a given social network during a pre-defined period of time and produce their specific analysis or predictions. The availability of open source research platforms in this area is scarse. Usually researchers use specific APIs and software modules to produce their studies. One research platform is Trendminer \cite{systems1}, an open source platform for real time analysis of Twitter data. It is the most similar work to POPmine although it only focus on Twitter and does not include sentiment analysis module, while POPmine allows cross media data collection and analysis, including sentiment.

\section{Architecture Overview}

Designing a system for tracking political opinion on the web such as POPmine must follow a set of technical and operational requirements:

\begin{itemize}

\item \textbf{Batch and real-time operation:} such a system must naturally be able to operate in real-time, i.e. collecting data as it is generated, processing it and updating indicators. However, it is also important to be able to operate in batch mode, in which it collects specific data from a period indicated by the user, if available, and then processes it. The system should use a distributed approach to deal with great volumes of data, (e.g. Hadoop). It should also be able to operate autonomously for long periods of time, measured in months.

\item \textbf{Adaptability:} the system should be able to adapt its models (e.g. sentiment classification) through time as well as across different applications. Updating models often requires manually annotated data (e.g. sentiment classification). Therefore the system should provide a flexible annotation interface.

\item \textbf{Modularity:} researchers should be able to plug in specific modules, such as a new data source and respective crawler or a different visualization. POPmine interfaces should use REST APIs and JSON data format, which allow users to add new modules that interact with other data sources (e.g. Wikipedia or Facebook). 

\item \textbf{Reusability:} the system should allow repeatability of all experiments to allow the research community to obtain equal results. We will make the software package publicly available as well as the data sources and configuration parameters used in experiments.

\item \textbf{Language independence:} each component of the system should apply a statistical language modeling completely agnostic to the language of the texts.

\end{itemize}

\begin{figure}[H]
    \centering
    \includegraphics[width=0.47\textwidth]{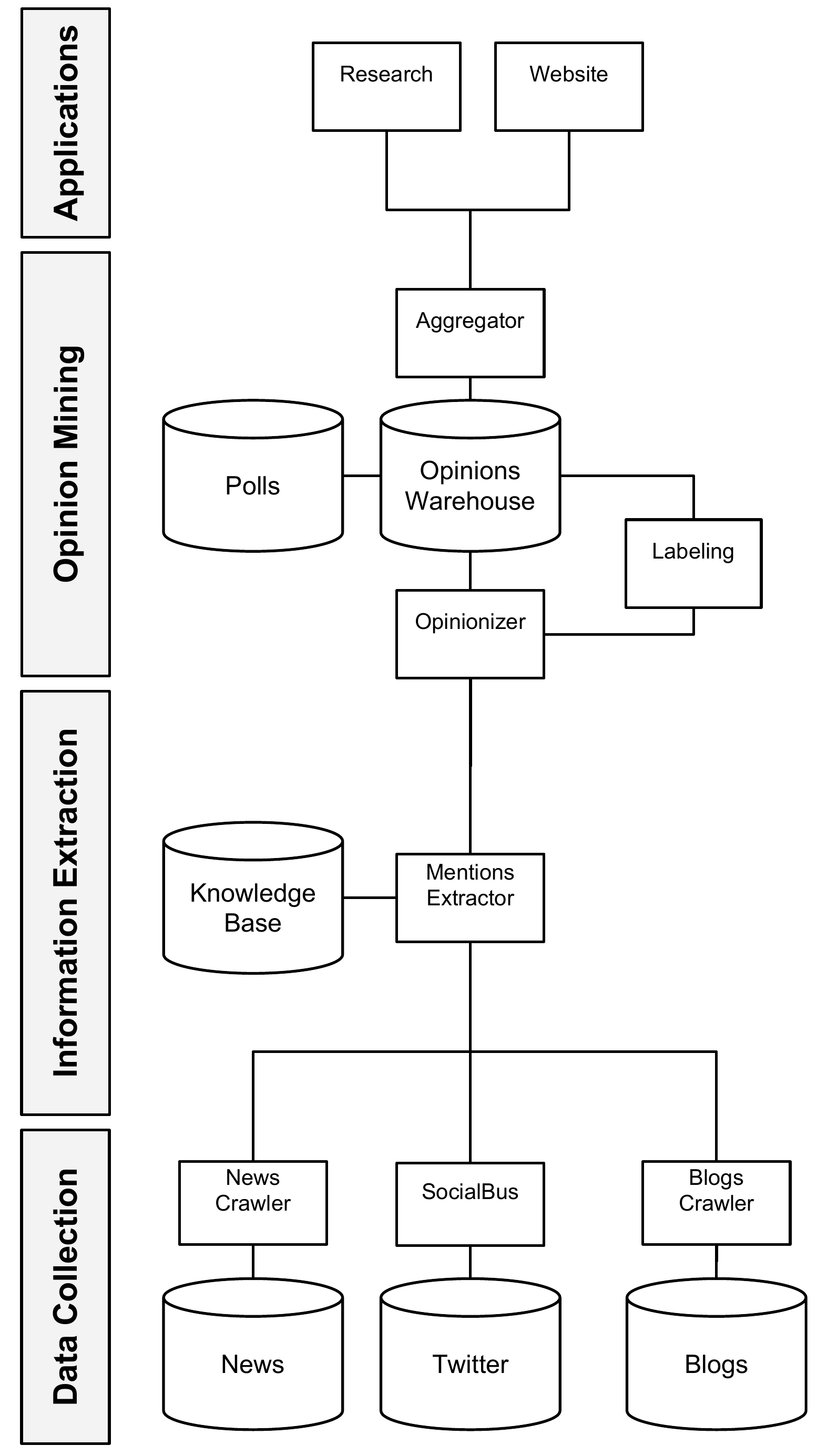}
    \caption{POPmine Architecture}
    \label{fig:architecture}
\end{figure}

To comply with the requirements we opted to create a specific architecture as depicted in Figure \ref{fig:architecture}. We present four major components: data collection, information extraction, opinion mining and applications.

We opted to use data from news articles, tweets and blog posts and each of these data sources requires its specific crawler. News articles and blog posts are collected using RSS feeds which eases the implementations of a specific crawler. Collecting data from Twitter poses some challenges. The need for large amounts of data, coupled with Twitter’s imposed limits demand for a distributed system. We opted to SocialBus\footnote{\url{http://reaction.fe.up.pt/socialbus/}}
 which enables researchers to continuously collect data from particular user communities, while respecting Twitter’s imposed limits.

The information extraction component comprises a knowledge base containing meta-data about entities, e.g., names or jobs. Using a knowledge base is crucial to filter relevant data mentioning politicians, such as news, tweets and blog posts. In our application scenario, we opted to use Verbetes, a knowledge base which comprises names, alternative names, and professions of portuguese people mentioned often in news articles.  We  explain the module for mentions extraction in Section V. 

The opinions warehouse contains the messages filtered by the information extraction component and applies a sentiment classifier to those messages using the Opinionizer classifier \citep{amirSemEval2014} as described in Section VI. One of the requirements of the Opinionizer is to use manually labeled data to train the classifier. We developed an online annotation tool for that effect.

We create opinion and polls indicators using the aggregator which is responsible to apply aggregation functions and smoothing techniques as described in Section VII. Once we obtain the aggregated data we make available a set of web services that can be consumed by different applications such as the POPSTAR website or other research experiences, such as polls predictions using social media opinions.

\section{Data Collection}

The data collection components crawl data from specific data sources which implement specific web interfaces (e.g. RSS feeds, Twitter API). Each data source must have its own data collection module which in turn connects to the POPmine system using REST services. POPmine stores data collected in a document oriented NoSQL database (MongoDB). This configuration allows modularity and flexibility, allowing the possibility of developing specific data collection components tailored to specific data sources. 

The default setting of data collection modules comprise the following components:

\begin{itemize}

\item \textbf{News:} Data from online news are provided by the service Verbetes e Notícias from Labs Sapo. This service handles online news from over 60 Portuguese news sources and is able to recognize entities mentioned in the news.

\item \textbf{Blogs:} Blog posts are provided by the blogs’ monitoring system from Labs Sapo, which includes all blogs with domain sapo.pt, blogspot.pt (Blogger) and Wordpress (blogs written in Portuguese).

\item \textbf{Twitter:} Tweets are collected using the platform SocialBus, responsible for the compilation of messages from 100.000 Portuguese users of Twitter. Tweets are collected in real time and submitted to a language classification. In our experiments we opted to collect the tweets written in Portuguese.

\end{itemize}

\section{Information Extraction}

The Information Extraction components address two tasks: Named Entity Recognition and Named Entitiy Disambiguation. We envision an application scenario where we need to track political entities. Usually this type of entities are well known therefore we opted to use a knowledge base to provide meta-data about the target entities, namely the most common surface forms of their names. Once we had the list of surface forms to search for we applied a sequential classification approach using a prefix tree to detect mentions. This method is very effective on news articles and blog posts but can result in noisy mentions when applied to Twitter. For instance, a tweet containing the word ``Cameron'' can be related with several entities, such as the UK prime minister, a filmmaker or a company. Furthermore, tweets are short which results in a reduced context for entity disambiguation. 

When monitoring the opinion of a given entity on Twitter, it is first necessary to guarantee that all tweets are relevant to that entity. Consequently, other processing tasks, such as sentiment analysis will benefit from filtering out noise in the data stream. The task we tackle consists in building a disambiguation classifier: given an entity $e_{i}$ and a tweet $t_{j}$  we want to classify $t_{j}$ as \emph{Related} or \emph{Unrelated} to $e_{i}$. 

We studied a large set of features that can be generated to describe the relationship between an entity and a tweet and different learning algorithms. Concerning features, we used tweets represented with TF-IDF, similarity between tweets and Wikipedia pages of target entities, Freebase entities disambiguation, feature selection of terms based on frequency and transformation of content representation using Singular Value Decomposition. The algorithms tested include Naive Bayes, Support Vector Machines, Random Forests, Decision trees and Neural networks.

The resulting classifier obtained the first place in the Filtering task of RepLab 2013 \citep{saleiro2013popstar}. The corpus used at the competition consisted of a collection of 90,000 tweets both in English and Spanish, possibly relevant to 61 entities.

\section{Sentiment Analysis}
We built a sentiment analysis module to detect and classify opinionated tweets, mentioning at least one political entity, as expressing a positive, negative or neutral opinion. This module was built with a supervised text classification approach based on the bag-of-words assumption, that is, ignoring word order and representing messages as high-dimensional feature vectors with the size of the vocabulary. Then, a manually labeled dataset was used to train a linear classifier that estimates the probability of a sentiment label, given a message. 

One of the main challenges of developing text classification systems for social media, is dealing with the small size of the messages and the large vocabularies needed to account for the significant lexical variation caused by the informal nature of the text. This causes feature vectors to become very sparse, hampering the generalization of the classifiers. Therefore, we first normalized the messages and then enriched the bag-of-words model with additional features based on dense word representations. These word representations, were derived from an unlabeled corpus of 10 Million tweets using two well-known unsupervised feature learning methods: (i) Brown \cite{brown1992class} hierarchical clustering algorithm, that groups together words that tend to appear in the same contexts. Words are thus represented as the cluster to which they belong; (ii) Mikolov \cite{mikolov2013a} neural language models that induce dense word vectors (commonly referred to as \textit{word embeddings}) by training a neural network to maximize the probability that words within a given window size are predicted correctly.

To capture the use of affective language, we extracted features that take into account the occurrence of words with known prior sentiment polarity, such as \textit{happy} or \textit{sad}. To this end, we considered terms from \textit{SentiLex-PT} \citep{Silva2012}, a Portuguese sentiment lexicon. In summary, each message was represented with a concatenation of the following features:

\begin{itemize}
\item bag-of-words (unigrams);
\item bag-of-words (word clusters);
\item sum of the word embeddings;
\item lexicon-based features;
\end{itemize}

Regarding the choice of the classifier, we evaluated several linear models available in the \emph{scikit-learn}\footnote{\scriptsize{\url{http://scikit-learn.org/}}} python library, attaining the best results with the L2-regularized Maximum Entropy model. The proposed system was validated in the context of two international shared task evaluations: \textit{RepLab} \citep{amigo2013overview,filgueiraspopstar} and \textit{SemEval: Twitter Sentiment Analysis} \citep{nakov2013,amirSemEval2014}, and ranked among the top places in both, attesting the adequacy of the approach. 

\section{Opinion Indicators}
\label{indicators}

\subsection{Buzz}

Buzz is the daily frequency with which political leaders are mentioned by Twitter users, bloggers and online media news. We use two types of indicators. The first type is the relative frequency with which party leaders are mentioned by each medium (Twitter, Blogs and News), on each day. This indicator is expressed, for each leader of each party, as a percentage relative to the total number of mentions to all party leaders. The second indicator is the absolute frequency of mentions, a simple count of citations for each political leader. 

To estimate trends in Buzz, we use the Kalman Filter. We allow users to choose the smoothing degree for each estimated trend. Users can choose between three alternatives: a fairly reactive one, where trend is highly volatile, allowing close monitoring of day-by-day variations; a very smooth one, ideal to capture long term trends; and an intermediate option, displayed by default.

\subsection{Sentiment} After identifying the polarity in each of the tweets,  there are several ways to quantify the overall sentiment regarding political leaders. We can, for instance, look at each target independently or in relative terms, compare positive with negative references or simply look at one side of the polarity, or look at daily, weekly or monthly data records.

In this first prototype we opted to present two separate indicators and their evolution across time, using in both cases the day as reference period. The fist indicator is the logarithm of the ratio of positive and negative tweets by political leader (party leaders and the president). In other words, a positive sign means that the political leader under consideration received more positive than negative tweets that day, while a negative result means that he received more negative than positive tweets. In mathematical notation:

The second approach is to simply look at the negative tweets (the vast majority of tweets in our base classifier) and calculate their relative frequency for each leader. In this way it is possible to follow each day which party leaders were, in relative terms, more or less subject to tweets with negative polarity. In mathematical notation:

\[ negatives share = \frac{negatives_{i,d}}{\sum_{negatives_{d}}} \]

\[ log sentiment_i =  log (\frac{positives_i + 1}{negatives_i +1})\]

\section{Application}

\begin{figure}[h]
    \centering
    \includegraphics[width=0.47\textwidth]{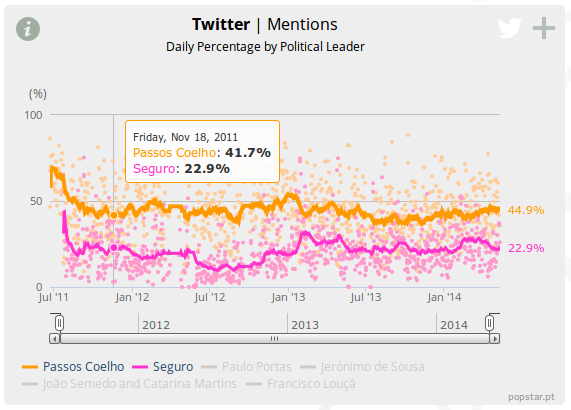}
    \caption{Twitter buzz share of political leaders.}
    \label{fig:site_buzz_share}
\end{figure}

We created a website\footnote{\url{http://www.popstar.pt}} to allow interactive visualization of the data collected and processed in real time by the POPmine platform. The site was developed within the scope of the POPSTAR project (Public Opinion and Sentiment Tracking, Analysis, and Research) and presents the following data: a) mentions to Portuguese party leaders in Twitter, in the blogosphere and in online news; b) sentiment conveyed through tweets regarding party leaders, c) voting intentions for the main political parties, measured by traditional polls; and d) evaluation of the performance of said party leaders, measured by polls. An example chart is depicted in Figure \ref{fig:site_buzz_share}.

Besides providing our indicators in the form of charts, the website also has a dashboard offering a more compact view of trends across indicators for all politicians.

\section{Conclusions}

We present an open source system capable of tracking political opinion on the web. The system implementation followed a set of requirements, namely batch and real-time operation, adaptability, modularity, reusability and language independence. As far as we know, this is the first open source plataform that integrates data collection, information extraction, opinion mining and visualization of political opinion data.

The system is able to collect texts from web-based conventional media (news items in mainstream media sites) and social media (blogs and Twitter) and to process those texts, recognizing topics and political actors, analyzing relevant linguistic units, and generating indicators of both frequency of mention and polarity (positivity/negativity) of mentions to political actors across sources, types of sources, and across time.

We hope to extend the system to allow topic classification of messages, such as economy, health or education. The system will also allow entity based retrieval as well as integration of open information extraction methods to depict relations between entities.

\section*{Acknowledgments}

This work was partially supported by FCT (Portuguese research funding agency) under project grant PTDC/CPJ-CPO/116888/2010 (POPSTAR), through contracts Pest-OE/EEI/LA0021/2013, EXCL/EEI-ESS/0257/2012 (DataStorm) and Ph.D. scholarship SFRH/BD/89020/2012. This research was also funded by SAPO Labs.

\bibliographystyle{unsrt}
\bibliography{refs}  

\end{document}